\documentclass{article}
\usepackage{graphicx}
\usepackage{latexsym}
\newcommand{\be}{\begin{equation}}
\newcommand{\ee}{\end{equation}}
\newcommand{\bea}{\begin{eqnarray}}
\newcommand{\eea}{\end{eqnarray}}

\begin{document}

\title{\bf Quasi-local Casimir energy and vacuum buoyancy in a weak gravitational field}

\author{
Francesco Sorge\\ 
{\it I.N.F.N. - Compl. Univ. di Monte S. Angelo,}\\
{\it via Cintia, Edificio 6,  I-80126 Napoli, Italy}\\
{\rm sorge@na.infn.it}}

\maketitle

\begin{abstract}
Casimir energy in presence of a weak gravitational field is discussed taking into account the issues related to energy and its conservation in a curved background. It is well-known that there are inherent difficulties in defining energy in General Relativity, essentially due to its non-localizability.
Using the concept of quasi-local mass and energy, it is shown that it is possible to attribute a Tolman mass to a massless scalar field confined to a Casimir cavity. Such non-local mass coincides - as expected - with the  Casimir energy.  The present approach offers an alternative derivation of the vacuum buoyancy force acting on a Casimir cavity, confirming the results presented by Calloni {\em et al.} in a series of papers devoted to explore the possibility of experimentally weighting the Casimir vacuum  (the so-called {\em Archimedes Experiment}).
  
\end{abstract}

pacs: 04.20.-q, 04.62.+v


\section{Introduction}
It is well-known that there are inherent difficulties in defining energy in General Relativity, essentially due to its non-localizability. Actually, a local density of energy related to the gravitational field cannot be defined as a consequence of the Equivalence Principle. Differently stated, a freely falling point-like particle doesn't experience any gravitational force. On the other hand, this is not the case when an extended body is taken into account.

It was Einstein \cite{einstein} who originally pointed out that energy and momentum conservation laws required to involve a gravitational energy tensor. The latter, being not covariant under general coordinate transformations, causes conservation laws to be obeyed only in some peculiar coordinate systems and reference frames \cite{landau,MTW}.

A possible way out to avoid such issue is to define a conserved quasi-local energy in the asymptotic (flat) regions of the spacetime. Indeed, such approach leads to envisage several kinds of quasi-local quantities (see, e.g., the comprehensive review by Szabados \cite{gorg}). Among them, the Tolman mass \cite{tolman1,tolman2,whittaker} is one of the standard definitions of quasi-local mass, other examples being, e.g., the ADM mass \cite{ADM} and the Komar mass \cite{komar}. All these definitions involve the Einstein equations, so that the resulting mass happens to play the active role of {\em gravitational} mass.
 

On the other hand, the debate  concerning the definition of the {\em inertial} mass-energy of an extended body in presence of gravitation has puzzled physicists since the birth of General Relativity, the latter being rooted in the Equivalence Principle, according to which inertial and gravitational mass should coincide  (see, e.g., \cite{ohanian,ram}).

Such intriguing matter unavoidably involves also quantum systems, which intrinsically suffer non-localizability \cite{sonego,onofrio}, because of the Heisenberg uncertainty principle.

In that respect,  the Casimir effect \cite{casimir1,casimir2}, represents an interesting arena for a theoretical analysis of the influence of a gravitational field upon an extended quantum system. Through the years, Casimir vacuum energy in presence of gravitation has been widely investigated and discussed in the literature \cite{setare,calloni1,calloni2,sorge1,bimonte2,bimonte3,esposito,bimonte1,sorge2,nazari0,sorge3,bezerra,nazari,zhang,sorge4,sorgej,sorgejj,sorge5,lima1,lima2}. A review about the topic can be found, e.g., in \cite{bordag,dalvit}. 

The issue whether vacuum fluctuations gravitate or not, has been deeply addressed in a series of papers by Milton {\em et al.}  \cite{milton1,milton2,milton3,milton4}. Also, the equivalence between gravitational and inertial mass of Casimir energy has been discussed, e.g., in \cite{milton5}. 

On the experimental side, Calloni {\em et al.} \cite{garufi,avino} have put forward the challenging proposal of weighting the quantum vacuum in a Casimir apparatus - the so-called {\em Archimedes’ Experiment}.

In an idealized scenario, a small Casimir cavity, suspended in the laboratory, is weighted in order to measure the contribution due to the vacuum expectation value of the mass-energy of a massless quantum field confined to the cavity.

As Calloni pointed out \cite{garufi}, when considering the forces applied to a stressed body, it is generally expected that {\em also} the spatial components of the stress-energy tensor contribute to the body mass, just as in the above recalled definitions of quasi-local mass. In particular, this should be the case of a suspended Casimir apparatus, kept at rest against the gravitational force acting upon it - its weight.

Considering a suspended Casimir cavity, Calloni proved that the net force required to balance the weight of the apparatus can be obtained properly red-shifting the force densities acting on the various points of the apparatus itself, in full agreement with previous results by Fulling {\em et al.} \cite{milton1} and  by Bimonte {\em et al.} \cite{bimonte3}. As a consequence, the mass of the apparatus is independent of the spatial sector of the stress-energy tensor, only depending on the $T^t_{\,\,\,t}$ component. The main result of such analysis is the existence of a tiny force, directed upwards and corresponding to the {\em weight} of the vacuum modes not allowed inside the cavity, due to the quantum field confinement. As suggested by Calloni, such force can be considered as an Archimedes buoyancy force in vacuum.

In the present paper we will show that the same results can be obtained applying the notion of (quasi-local) Tolman mass to a Casimir apparatus. We will prove that, due to a subtle cancellation between the spatial sectors of the stress-energy tensors pertaining to the cavity walls and the confined quantum field, only  the  $T^t_{\,\,\,t}$ components do matter. Such a cancellation is likely to be traced back to the virial theorem \cite{landau}.

In particular, we will confirm the existence of a  buoyancy force due to the interaction between the Casimir vacuum and the weak gravitational field experienced by the quantum field confined to the cavity.  

A key-point is that the mass-energy of the apparatus does depend on the reference frame. This partly helps understanding early contrasting claims in the literature, concerning the influence of a weak gravitational field on the Casimir energy. On the contrary, the implied gravitational interaction appears to be {\em frame-independent}, hence physically observable, owing to the gauge-invariance of the coupling between matter and gravitational field.

The paper is organized as follows. In section 2 we briefly recall the standard approach to the evaluation of Casimir energy in a flat space-time.  In section 3 we discuss the issue related to energy definition in presence of gravity, recalling the notion of quasi-local energy. In section 4 we investigate the relationship between the quasi-local Tolman mass and the energy content of a Casimir apparatus in a weak gravitational field. In section 5 we introduce the Fermi coordinates adapted to a reference frame in which a Casimir cavity is kept at rest against the gravitational field of a weak static source (like the Earth). In section 6 we define a Tolman mass related to the Casimir quantum vacuum. We subsequently prove that such mass indeed matches the Casimir energy. This result can be traced back to the virial theorem.  In section 7 we propose a lagrangian approach to the evaluation of the stress-energy tensor pertaining to a massless scalar field confined to the Casimir cavity in presence of a weak gravitational field. Finally, in section 8 the stress-energy tensor is employed to evaluate the mass-energy contribution of the quantum vacuum.
Section 9 is devoted to some concluding remarks.  

Throughout the paper, unless otherwise specified, use has been made of natural units:  $c=1$, $\hbar=1$, $G=1$. Greek indices take values from 0 to 3; latin ones take values from 1 to 3. The metric signature is $-2$, with determinant $g$.


\section{Casimir energy}
In this section we will briefly recall the flat-spacetime standard approach  to the evaluation of Casimir energy for a massless scalar field $\phi(x)$ confined to a small cavity, delimited by two parallel plates, of  area $A_\perp$, perfectly reflecting and placed at a distance $L$ each other. 
We will also assume that the plate separation satisifies the condition $L\ll\sqrt{A_\perp}$, so that edge effects can be neglected.

In a Minkowski spacetime, and assuming a coordinate set $\{t,\,x,\,y,\,z\}$ with the $z$ axis orthogonal to the cavity plates, the confinement of the scalar field  requires the field modes $\phi(x)$ to satisfy the Klein-Gordon equation
\be\label{kg}
\eta^{\mu\nu}\partial_\mu\partial_\nu\phi(x)\equiv\Box\phi(x)=0,
\ee
together with the Dirichlet boundary conditions at the cavity walls
\be\label{dirichlet}
\phi(x) |_{z=0}=\phi(x) |_{z=L}=0.
\ee
The field eigenmodes $\phi_n(x)$, normalized according to the Klein-Gordon scalar product \cite{birrell}
\be\label{KGscalprod}
\langle \phi_1(x),\phi_2(x)\rangle=-i\int_Vd^3x\big[\phi_1(x)\partial_t\phi_2^*(x)-\left(\partial_t\phi_1(x)\right)\phi_2^*(x)\big],
\ee
(with $V$ being the cavity volume), are found to be
\be\label{phin}
\phi_{k,n}(x)=\frac{1}{2\pi (L\omega_{k,n})^{1/2}}e^{-i\omega_{k,n} t}e^{i\vec k_\perp\cdot\vec x_\perp}\sin\bigg(\frac{n\pi}{L}z\bigg),
\ee
where $\vec k_\perp=(k_x,\,k_y)$ and $\vec x_\perp =(x,\,y)$.
The corresponding eigenfrequencies $\omega_{k,n}$ are
\be\label{omegan}
\omega_{k,n}^2=k_\perp^2+\bigg(\frac{n\pi}{L}\bigg)^2.
\ee
The Casimir energy is then evaluated as the vacuum expectation value 
\be\label{ecas}
E=\int_V d^3x\langle 0|T^t_{\,\,t}|0\rangle=\int_V d^3x\sum_n\int d^2k_\perp T^t_{\,\,t}[\phi_{k,n}(x),\phi_{k,n}^*(x)],
\ee
where
\be\label{bilin}
 T_{tt}[\phi_{k,n}(x),\phi_{k,n}^*(x)]=\partial_t\phi_{k,n}\partial_t\phi_{k,n}^*-\frac{1}{2}\eta_{tt}\eta^{\alpha\beta}\partial_\alpha\phi_{k,n}\partial_\beta\phi_{k,n}^*.
 \ee
 Substituting (\ref{phin}) in (\ref{ecas}) we obtain
 \be
 E=A_\perp\sum_n\int d^2k_\perp\int_0^L dz \frac{1}{(2\pi)^2L\omega_{k,n}}\bigg[k_\perp^2\sin^2\bigg(\frac{n\pi}{L}z\bigg)+\frac{1}{2}\bigg(\frac{n\pi}{L}\bigg)^2\bigg].
 \ee
 Integration over the cavity volume $V=A_\perp L$ yields
 \be\label{Ediv}
 E=A_\perp \frac{1}{2(2\pi)^2}\sum_n\int d^2k_\perp \bigg[k_\perp^2+\bigg(\frac{n\pi}{L}\bigg)^2\bigg]^{1/2}.
 \ee
 The divergent integral can be managed by means of dimensional regularization, using
 \be\label{dimreg}
 \int d^2k_\perp\frac{1}{(k_\perp^2+s)^\alpha}=\pi\frac{\Gamma(\alpha-1)}{\Gamma(\alpha)}\frac{1}{s^{\alpha-1}},
 \ee
 thus obtaining
 \be
  E=-A_\perp \frac{\pi^2}{12 L^3}\sum_n n^{3}.
  \ee
  The infinite sum can be carried on by means of the analytic continuation of the involved Riemann zeta function \cite{edwards,elizalde}
  \be
  \zeta(-3)=\frac{1}{120}.
  \ee
  We finally get the well-known result for the Casimir energy
  \be\label{ecasflat}
 E=-A_\perp \frac{\pi^2}{1440 L^3}.
 \ee


\section{Quasi-local energy}
In a curved, albeit static spacetime, equation (\ref{ecas}) is usually modified taking into account the projection on the local static observer  $u^\mu=\delta^\mu_t/\sqrt{g_{tt}}$, hence obtaining
\be\label{Ecurved}
E=\int_\Sigma d^3x\,\sqrt{g_\Sigma}\, u^\mu u^\nu \langle 0|T_{\mu\nu}|0\rangle=\int_\Sigma d^3x\,\sqrt{g_\Sigma}\, g^{tt} \langle 0|T_{tt}|0\rangle,
\ee
where $g_\Sigma$ is the determinant of the induced metric on $\Sigma$ (the cavity volume).
However, (\ref{Ecurved}) has not a well defined physical meaning, eventually being a mere summation of the vacuum expectation energy values as measured by different static observers placed at different points of the three-dimensional cavity domain $\Sigma$. Those observers experience different time flows, so they are physically not-equivalent. Hence (\ref{Ecurved}) does not reflect any realistic measurement performed in a suitable experiment.

As recalled in the Introduction, similar difficulties in defining  a local energy-momentum density in a curved spacetime lead to consider  energy and angular momentum as non-local observables  for extended domains of spacetime.
Here we will briefly review the standard approach \cite{moller,wald,fdf}, considering - for sake of simplicity - a {\em classical} system. In section 8 the present technique will be straightforwardly applied to the {\em semi-classical} case involving a quantum system - namely a Casimir apparatus.

Let  $({\cal M}, {\bf g})$ be a {\em static} spacetime, admitting a foliation, equipped with the metric
\be\label{gstatic}
ds^2=g_{tt}dt^2+g_{ij}dx^idx^j.
\ee
Let $\Omega\subset {\cal M}$ be a domain of compact closure, confined to a world-tube $W$, delimited by a time-like surface $S$ and two space-like hypersurfaces $\Sigma_1$ and $\Sigma_2$, being the latter  in the future of the former ($t_2>t_1$). Provided the induced metric on $\partial\Omega$ (the boundary of $\Omega$) is non-degenerate, the three-dimensional hypersurface element on $\Sigma$ reads $d^3\Sigma_\mu=n_\mu d\Sigma$, where $d\Sigma=\sqrt{g_\Sigma}d^3x$ is the invariant volume element on $\partial\Omega$, $g_\Sigma$ is the determinant of the induced metric and $n^\mu$ is a unit vector normal to the hypersurface.

Consider now a physical system whose stress-energy tensor $T^{\mu\nu}$ obeys the following requirements:
\begin{itemize}
\item[a)]  $T^{\mu\nu}$ is vanishing outside the domain $\Omega$;
\item[b)] $\nabla_\mu T^\mu_{\,\,\,\nu}=f^{ext}_{\nu}$, where $f^{ext}_{\nu}$, describes any non-gravitational external force density acting upon the system (in other words, we allow for a {\em non-closed} system) \cite{moller,dalvit};
\item[c)] $T^{\mu\nu}$ does not appreciably influence the static spacetime background.
\end{itemize}
Due to (c), the background is (almost) static\footnote{The quantum fluctuations inside the cavity evolve on time-scales quite smaller than the ones involved during typical laboratory measurements. Hence, the {\em static} spacetime assumption is safely preserved.}, so it admits a time-like Killing vector $\vec\xi\equiv\vec\partial_t=\delta^\mu_t\vec\partial_\mu$. Consider then the following quantity
\be\label{Q}
Q=\int_\Omega d\Omega\,\nabla_\mu\left(\xi_\nu T^{\mu\nu}\right),
\ee
which can be converted to a hypersurface  integral
\bea\label{Qsurface}
Q=\oint_{\partial\Omega}d\Sigma\, n_\mu\left(\xi_\nu T^{\mu\nu}\right)\nonumber\\
=\int_{\Sigma_2\cap\Omega}d\Sigma\, n_\mu\xi_\nu T^{\mu\nu}-\int_{\Sigma_1\cap\Omega}d\Sigma\, n_\mu\xi_\nu T^{\mu\nu}.
\eea
Recalling  (\ref{gstatic}), using $\xi^\mu=\delta^\mu_t$ and $n^\mu=\delta^\mu_t/\sqrt{g_{tt}}$, we write
\be\label{Qsurface2}
Q=\int_{\Sigma_2\cap\Omega}d^3x\sqrt{-g} T^t_{\,\,\,t}-\int_{\Sigma_1\cap\Omega}d^3x\sqrt{-g} T^t_{\,\,\,t}.
\ee 
On the other hand, from (\ref{Q}) we have
\be\label{Qalt}
Q=\int_\Omega d\Omega\,\left(\nabla_\mu\xi_\nu\right) T^{\mu\nu}+\int_\Omega d\Omega\,\xi_\nu\nabla_\mu T^{\mu\nu}.
\ee
The first term vanishes, owing to the Killing equations $\nabla_{(\mu}\xi_{\nu)}=0$. Recalling also (\ref{gstatic}) and using $\xi^\mu=\delta^\mu_t$ again, we find
\be\label{Qalt2}
Q=\int_\Omega d\Omega\,\xi_\nu\nabla_\mu T^{\mu\nu}=\int_\Omega d\Omega\,\nabla_\mu T^{\mu}_{\,\,\,t}=\int_\Omega d^4x\sqrt{-g}\,f^{ext}_t,
\ee
where use has been made of assumption (b). Notice that $f^{ext}_t$ can be considered as the (non-gravitational) exchanged power density. Comparing (\ref{Qsurface2}) and (\ref{Qalt2}) we have
\be\label{deltaE}
\int_{\Sigma_2\cap\Omega}d^3x\sqrt{-g} T^t_{\,\,\,t}-\int_{\Sigma_1\cap\Omega}d^3x\sqrt{-g} T^t_{\,\,\,t}=\int_\Omega d^4x\sqrt{-g}\,f^{ext}_t.
\ee
The RHS can now be interpreted as the total amount of energy exchanged between the system and the surroundings during the time lapse $t_2-t_1$. Hence the the LHS can be considered as the variation in the amount of energy inside the system. We can then define 
\be\label{NLE}
E_t=\int_{\Sigma_t\cap\Omega}d^3x\sqrt{-g} T^t_{\,\,\,t},
\ee
the energy of the system at time $t$. 

Since $\sqrt{-g}=\sqrt{g_{tt}}\sqrt{g_\Sigma}$, 
comparing (\ref{NLE}) to (\ref{Ecurved}) we see that the difference amounts to an extra factor $\sqrt{g_{tt}}$, thus accounting for the {\em red-shift} suffered by the locally measured energy density when referred to a distant (ideally placed at the spatial infinity) inertial observer.

Notice that, in the case of an {\em isolated} system, $f^{ext}_t=0$ and energy is {\em conserved}.  This is indeed an example of conserved {\em charge} in General Relativity (see, e.g. \cite{aoki}).


\section{Relationship with Quasi-local Tolman mass}
The integration measure $d^3x\sqrt{-g}$ appearing in (\ref{NLE}) is reminiscent of that contained in the definition of the Tolman mass \cite{tolman1,tolman2,whittaker}. Tolman proved that, provided the solution of Einstein equations is time-independent and free of singularities, the mass of the source can be given as a volume integral over the diagonal components of the source stress-energy tensor, namely 
\be\label{Tm}
M=\int_Vd^3x\sqrt{-g}\left(T^t_{\,\,\,t}-T^x_{\,\,\,x}-T^y_{\,\,\,y}-T^z_{\,\,\,z}\right).
\ee
We stress that, owing to (\ref{Tm}),  in general {\em all} the diagonal components of the stress-energy tensor (not only the $T^t_{\,\,\,t}$ component) contribute to the mass. 

In a static spacetime we can use (\ref{Tm}) to describe  the (quasi-local) mass of a physical system whose stess-energy tensor is $T^{\mu\nu}$. Although in (\ref{Tm}) the metric is determined by  $T^{\mu\nu}$ through the Einstein equations, the above mass definition may apply {\em also} when the spacetime background is {\em fixed}, i.e.,  the stress-energy tensor of the physical system we are considering has negligible influence on the background. 

Let us show this in more detail. Consider the case we are interested in, namely a small Casimir apparatus suspended in a laboratory at the Earth surface, in a weak gravitational field. The Casimir apparatus is an example of {\em non-closed} system, since an external force is required to keep it at rest against the gravitational field of the Earth. However, we may think of the composite system Earth + Casimir apparatus (E+C), as an (almost) {\em isolated} system,  obeying the continuity equation 
\be\label{SETtotal}
\nabla_\mu T^{\mu\nu}=\nabla_\mu\left(T_{(E)}^{\mu\nu}+T_{(C)}^{\mu\nu}\right)=0,
\ee
where $T_{(E)}^{\mu\nu}$ and $T_{(C)}^{\mu\nu}$ represent the stress-energy tensors of the Earth and of the Casimir apparatus, respectively. 
The corresponding Tolman mass of the composite system E+C is
\be\label{MEC}
M=\int_{V_\infty}d^3x\,\sqrt{-g_T}\big[T^t_{\,\,\,t}-T^i_{\,\,\,i}\big],
\ee
where $g_T$ is the determinant of the metric due to the {\em whole} E+C system and $V_\infty$ refers to an integration volume large enough to guarantee $T^{\mu\nu}\sim 0$ on the boundary $\partial V_\infty$ (for further details, see, e.g., de Felice \& Clarke \cite{fdf}).
The above integral splits into
\be\label{split}
M=\int_{V_\infty}d^3x\,\sqrt{-g_T}\big[T^{t}_{\,\,\,t}-T^i_{\,\,\,i}\big]_{(E)}+\int_{V_\infty}d^3x\,\sqrt{-g_T}\big[T^t_{\,\,\,t}-T^i_{\,\,\,i}\big]_{(C)},
\ee
with an obvious notation meaning.
Since $|T^{\mu\nu}_{(C)}|\ll|T^{\mu\nu}_{(E)}|$, the spacetime metric will basically coincide with the one due solely to the Earth. So $\sqrt{-g_T}\simeq\sqrt{-g}$, where $g$ is the determinant of that we named the {\em background} metric, whose gravitational source is the Earth mass only. Hence
\be\label{split2}
M=\int_{V_\infty}d^3x\,\sqrt{-g}\big[T^t_{\,\,\,t}-T^i_{\,\,\,i}\big]_{(E)}+\int_{V_\infty}d^3x\,\sqrt{-g}\big[T^t_{\,\,\,t}-T^i_{\,\,\,i}\big]_{(C)}.
\ee
The first term in the LHS of (\ref{split2}) represents the Tolman mass of the Earth, $M_E$. The last term can be considered as the quasi-local Tolman mass of the Casimir apparatus
\be\label{m}
m=\int_{V_\infty}d^3x\,\sqrt{-g}\big[T^t_{\,\,\,t}-T^i_{\,\,\,i}\big]_{(C)}.
\ee
Notice that in (\ref{m}) the implied metric is just that of the background, as previoulsy stated. We conclude the present section pointing out that in (\ref{m}) the integration volume can be reduced to the Casimir cavity volume $V$, since $T_{(C)}^{\mu\nu}=0$ outside the cavity. Thus we write
\be\label{m2}
m=\int_V d^3x\,\sqrt{-g}\big[T^t_{\,\,\,t}-T^i_{\,\,\,i}\big]_{(C)}.
\ee


\section{Fermi normal coordinates}
 
Consider now the case of a Casimir apparatus, placed at rest in a weak gravitational field - e.g., in a laboratory on the surface of the Earth. Let us introduce a Fermi reference frame \cite{MTW,manasse,fermi}. Recall that the Fermi frame represents the closest generalization to a weak gravitational field of an inertial coordinate system in flat space. 
As pointed out in \cite{bimonte2}, construction of these coordinates involves only invariant quantities such as the observer's proper time and  geodesic distances from the world-line. This, in turn, allows a clear identification of the various quantities appearing in the metric, also avoiding gauge issues.

To first order, and neglecting rotations, the metric of a weak static gravitational field, due to a spherically symmetric mass, a radial distance $z$ above a surface of radius $R$, reads 
\be\label{ds}
ds^2=(1+2\gamma z)dt^2-dx^2-dy^2-dz^2+O(|x|^2),
\ee
where $\gamma=M/R^2$ is the gravitational acceleration of the observer with respect to a locally freely falling observer.

Notice that (\ref{ds})  can be also deduced from the weak-field spacetime line element in the Lorentz gauge by means of a suitable coordinate transformation (see Appendix). The underlying physics is not altered by such transformation, owing to the gauge-invariance of the gravitational interaction term, as discussed in the Appendix.
We will suppose a  Casimir cavity, whose plates, of proper area $A_\perp$, are separated by a proper distance $L\ll\sqrt{A_\perp}$  and orthogonal to the gravitational acceleration $\gamma$. 

We point out that the weak field approximation can be safely employed in the present approach, since the typical cavity size ($\sim L$) is indeed much smaller than the Earth radius $R$. Hence, the gravitational acceleration $\gamma$ can be considered almost uniform throughout the whole Casimir apparatus.

We will also suppose that the lower plate is placed at a proper height $H$ with respect to the adopted Fermi reference frame, so that the plates are at $z=H$ and $z=H+L$, respectively.

Given the above assumptions, we will now move to evaluate the mass-energy content of the Casimir cavity.


\section{Tolman mass of Casimir energy}
The title of the present section seems somewhat odd, since in Relativity one is used to {\em identify} the concepts of mass and energy. In that respect, we would like to consider (\ref{NLE}) and (\ref{m2}) as {\em equivalent}. Nevertheless, the corresponding expressions are quite different. On the other hand it is well-known that - when considering stressed systems - the space-like components of the stress-energy tensor contribute as well to the mass of the system. As pointed out in the introduction, a Casimir apparatus is indeed a stressed body. So, what about such issue?

Quite interestingly, the solution is rooted in the mutual cancellation between the space-like components of the stress-energy tensor of the quantum field enclosed in the cavity {\em and} the stress-energy tensor of the cavity itself, as we will prove below.

Recall that the Casimir apparatus we are considering is a {\em non-closed system}, because of the external force required to keep it at rest against the gravitational field of the Earth. In the Earth-based laboratory frame, the required external force density $f^\nu$ satisfies
\be\label{equilibrium}
f_\nu+\nabla_\mu T^{\,\,\mu}_{(C)\,\nu}=0.
\ee
Expanding (\ref{equilibrium}) in the reference frame of the Fermi observer (\ref{ds}), we identify the $z-$component of the required force with the {\em weight} of the apparatus
\be\label{Fz}
F_z=\int_V d^3x\sqrt{-g}f_z=m\gamma=\gamma\int_V d^3x\sqrt{-g}\big[T^t_{\,\,\,t}-T^i_{\,\,\,i}\big]_{(C)},
\ee
where use has been made of (\ref{m2}). The equilibrium equation (\ref{equilibrium}) in the Fermi frame can be rewritten as follows
\be\label{equil2}
\gamma\int_V d^3x\sqrt{-g}\big[T^t_{\,\,\,t}-T^i_{\,\,\,i}\big]_{(C)}+\int_V d^3x\,\sqrt{-g}\bigg[\frac{1}{\sqrt{-g}}\partial_\mu T^{\,\,\mu}_{(C)\,z}-\frac{1}{2}\partial_z\left(g_{tt}\right)T_{(C)}^{tt}\bigg]=0.
\ee
Recalling that $g_{tt}=1+2\gamma z$ [see (\ref{ds})] , we get
\be\label{eq3}
\gamma\int_V d^3x\sqrt{-g}\big[T^t_{\,\,\,t}-T^i_{\,\,\,i}\big]_{(C)}+\int_V d^3x\,\partial_\mu T^{\,\,\mu}_{(C)\,z}-\gamma\int_V d^3x\sqrt{-g}T_{(C)}^{tt}=0.
\ee
The first and the last term cancel each other [at the $O(\gamma)$ order, $\gamma\,T_{(C)}^{tt}=\gamma\,T^{\,\,\,t}_{(C)\,t}$], so we are left with
\be\label{eq4}
-\gamma\int_V d^3x\sqrt{-g}T^{\,\,\,i}_{(C)\,i}+\int_V d^3x\,\partial_t T^{\,\,t}_{(C)\,z}+\int_V d^3x\,\partial_j T^{\,\,j}_{(C)\,z}=0.
\ee
The last term is zero, since it can be converted to a surface integral on the boundary $\partial V$ of the cavity volume, where $T_{(C)}^{\mu\nu}=0$. The second term is zero, as the cavity does not exchange energy and momentum with the surroundings. As a consequence
\be\label{eq5}
\int_V d^3x\sqrt{-g}T^{\,\,\,i}_{(C)\,i}=0.
\ee
Using this result in (\ref{m2}) we find
\be\label{m3}
m=\int_V d^3x\,\sqrt{-g}\,T^{\,\,t}_{(C)\,t}.
\ee
The stress-energy tensor of the Casimir apparatus can now be decomposed in a part related to the quantum field $\phi(x)$ confined to the cavity and a part containing the stress-energy tensor of the cavity itself:
\be\label{Tcomposed}
T_{(C)}^{\mu\nu}=T_{(\phi)}^{\mu\nu}+T_{(cav)}^{\mu\nu}.
\ee
Using (\ref{Tcomposed}) in (\ref{m3}) we get
\be\label{m4}
m=\int_V d^3x\,\sqrt{-g}\,T^{\,\,t}_{(\phi)\,t}+\int_V d^3x\,\sqrt{-g}\,T^{\,\,t}_{(cav)\,t}.
\ee
We easily identify the last term with the cavity mass, while the mass of the scalar field inside the cavity reads
\be\label{mphi}
m_\phi=\int_V d^3x\,\sqrt{-g}\,T^{\,\,t}_{(\phi)\,t}.
\ee
Comparing (\ref{mphi}) and (\ref{NLE}) we find that the two result agree, as expected. We finally understand the trick looking at (\ref{eq5}) where it appears evident the cancellation between the space-like components of the stress-energy tensors $T^{\mu\nu}_{(\phi)}$ and $T^{\mu\nu}_{(cav)}$. We would like to recall, in that respect, the solution of the well-known Tolman paradox, proposed by Misner and Putnam \cite{putnam}. We point out that such cancellation basically relies on the virial theorem \cite{landau} (see, e.g., Carlip \cite{carlip} and  Ohanian \cite{ohanian} for an exhaustive discussion; see also \cite{mitra1,mitra2,ehlers}).

We have thus proved that 
\begin{itemize}
\item[]{\em if}  (according to the Equivalence Principle) the gravitational and inertial (quasi-local Tolman) mass associated to the field quantum vacuum coincide, then the quasi-local mass value is related solely to the $T^{\,\,t}_{(\phi)\,t}$ component of the stress-energy tensor, as expected. The latter, in turn, carries information on the gravitational interaction with the confined quantum field.
\end{itemize}

Notice that the computed Tolman mass does depend on the nature of the enclosed quantum field. Different quantum fields (as, e.g., the electromagnetic one) imply different stress-energy tensors, hence different results for the Tolman mass.
Furthermore, the present approach was not intended to enforce the Equivalence Principle. On the contrary, it basically relies on the Equivalence Principle, which was indeed employed in deriving the result (\ref{mphi}).

The main result of the present section is that - although all the stress-energy tensor components play a role in determining the mass of a stressed Casimir apparatus - the effective contribution to the mass (energy) of the latter, due to the confined scalar field, can be evaluated simply starting  from the knowledge of the mixed time-time $T^{\,\,t}_{(\phi)\,t}$ component of the field stress-energy tensor.


\section{Weak gravity: a lagrangian approach}
In the present section we will provide a lagrangian derivation of the stress-energy tensor for a massless scalar field confined in a Casimir cavity, following a linearized gravity approach. Hereafter, we will suppress - for sake of clarity - the suffix $(\phi)$ appearing in $T^{\mu\nu}_{(\phi)}$, being understood that in what follows  the stress-energy tensor $T^{\mu\nu}$ refers to the quantum field $\phi(x)$.

The matter action for a massless scalar field $\phi$ in a curved spacetime background reads
\be\label{S}
S_m=\int d^4x\sqrt{-g}{\cal L},
\ee
where
\be\label{Lagr}
{\cal L}=\frac{1}{2}g^{\mu\nu}\nabla_\mu\phi\nabla_\nu\phi.
\ee
In the weak field limit $g_{\mu\nu}=\eta_{\mu\nu}+h_{\mu\nu}$, $|h_{\mu\nu}|\ll 1$,  we may expand the matter lagrangian as follows \cite{cetoli}:
\be\label{Lexpand}
{\cal L}_m=\sqrt{-g}{\cal L}=\big[\sqrt{-g}{\cal L}\big]_{h=0}+\bigg[\frac{\delta}{\delta g^{\mu\nu}}\big(\sqrt{-g}{\cal L}\big)\bigg]_{h=0}\delta g^{\mu\nu}+\cdots.
\ee
Since $g^{\mu\nu}=\eta^{\mu\nu}-h^{\mu\nu}$, we also have $\delta g^{\mu\nu}=g^{\mu\nu}-\eta^{\mu\nu}=-h^{\mu\nu}$. Furthermore, using the stress-energy tensor definition
\be\label{set}
T_{\mu\nu}=\frac{2}{\sqrt{-g}}\frac{\delta}{\delta g^{\mu\nu}}\big(\sqrt{-g}{\cal L}\big),
\ee
we rewrite (\ref{Lexpand}) up to the first order in $h$
\be\label{Lexpand2}
{\cal L}_m={\cal L}_0+{\cal L}_g,
\ee
where
\be\label{Lflat}
{\cal L}_0=\big[\sqrt{-g}{\cal L}\big]_{h=0}=\frac{1}{2}\eta^{\mu\nu}\partial_\mu\phi\partial_\nu\phi
\ee
is the flat spacetime field lagrangian and
\be\label{gravintlangr}
{\cal L}_g=-\frac{1}{2}T^{(0)}_{\mu\nu}h^{\mu\nu},
\ee
represents the contribution due to the gravitational interaction, with
\be\label{set0}
T^{(0)}_{\mu\nu}=\bigg[\frac{2}{\sqrt{-g}}\frac{\delta}{\delta g^{\mu\nu}}\big(\sqrt{-g}{\cal L}\big)\bigg]_{h=0}=\partial_\mu\phi\partial_\nu\phi-\eta_{\mu\nu}{\cal L}_0,
\ee
being the flat spacetime stress-energy tensor of the scalar field. Notice that the $\phi$'s appearing in (\ref{Lexpand2}) represent the solutions of the flat spacetime Klein-Gordon equation (\ref{kg}).
Combining (\ref{Lexpand2}), (\ref{Lflat}), (\ref{gravintlangr})  and (\ref{set0}) we also have 
\be\label{Lm}
{\cal L}_m=\bigg(1+\frac{1}{2}h\bigg){\cal L}_0-\frac{1}{2}h^{\mu\nu}\partial_\mu\phi\partial_\nu\phi,
\ee
where\footnote{Remember that index raising and lowering in the small quantities $h_{\mu\nu}$ is performed by means of the flat metric $\eta_{\mu\nu}$.} $h=h^\mu_{\,\,\mu}=\eta^{\mu\nu}h_{\mu\nu}$.

From ${\cal L}_m$ we readily obtain the stress-energy tensor for the scalar field interacting with the gravitational background\footnote{Notice that, at this stage, we cannot evaluate $T_{\mu\nu}$ using (\ref{set}) again, since now the background $g_{\mu\nu}$ is {\em fixed}.}. Using
\be\label{setgrav}
T^\mu_{\,\,\,\nu}=\frac{\partial{\cal L}_m}{\partial(\partial_\mu\phi)}\partial_\nu\phi-\delta^\mu_{\,\,\nu}{\cal L}_m,
\ee
we get
\be\label{setmixed}
 T^\mu_{\,\,\,\nu}=\bigg(1+\frac{1}{2}h\bigg)T^{(0)\mu}_{\quad\,\,\,\,\nu}-h^{\mu\beta}\partial_\beta\phi\partial_\nu\phi+\frac{1}{2}\delta^\mu_{\,\,\nu}h^{\alpha\beta}\partial_\alpha\phi\partial_\beta\phi.
\ee
For our purposes, we need the $T^t_{\,\,\,t}$ component in the adopted Fermi frame, in which the only non-null first-order metric correction to the minkowskian background is $h_{tt}=h^{tt}=h=2\gamma z$ [see (\ref{ds})]. Hence
\be\label{Tttmixed}
T^t_{\,\,\,t}=\left(1+\frac{1}{2}h\right)T^{(0)\,t}_{\quad\,\,\,\,t}-\frac{1}{2}h_{tt}\big(\partial_t\phi\big)^2.
\ee


\section{Quasi-local Casimir mass-energy}
We are now ready to evaluate the Casimir energy according to the Fermi observer. 
In the Fermi coordinates (\ref{ds}), $\sqrt{-g}=\sqrt{g_{tt}}\sqrt{g_\Sigma}=\sqrt{g_{tt}}$. 
Using (\ref{NLE}) and (\ref{Tttmixed}) we get 
\be\label{Ecas1}
E=\int_V\,d^3x\sqrt{-g}\langle T^t_{\,\,\,t}\rangle=\int_V\,d^3x\sqrt{g_{tt}}\langle T^t_{\,\,\,t}\rangle=A_\perp\int_H^{H+L}\,dz\langle T^{(0)t}_{\quad\,\,\,\,t}\rangle+\delta E.
\ee
Notice that the transition to the semi-classical framework has been formally made replacing $ T^t_{\,\,\,t}$ with $\langle T^t_{\,\,\,t}\rangle$, 
where  $\langle T^t_{\,\,\,t} \rangle\equiv \langle 0|  T^t_{\,\,\,t}|0\rangle$ [see (\ref{ecas})].

Recall that $H$ is the height at which the lower cavity plate is suspended and kept fixed with respect to the observer. The first term in the RHS of (\ref{Ecas1}) represents the flat spacetime result [see (\ref{ecas}) and (\ref{ecasflat})]. The correction $\delta E$ reads
\be\label{deltaE}
\delta E=A_\perp\int_H^{H+L}\,dz\bigg\langle \frac{1}{2}h_{tt}\big(\partial_i\phi\big)^2\bigg\rangle.
\ee
Using the field modes (\ref{phin}) we have
\bea\label{deltaEchange}
\delta E &=& A_\perp\sum_n\int d^2k_\perp N_n^2\int_H^{H+L} dz\, \gamma z\bigg[k^2_\perp\sin^2\bigg(\frac{n\pi}{L}(z-H)\bigg)\nonumber\\
&+&\bigg(\frac{n\pi}{L}\bigg)^2\cos^2\bigg(\frac{n\pi}{L}(z-H)\bigg)\bigg],
\eea
where [see (\ref{phin})]
\be\label{Nn}
N_n=\frac{1}{2\pi (L\omega_{k,n})^{1/2}}.
\ee
The replacement $z\rightarrow z-H$ in the field modes $\phi_{k,n}(x)$ appearing in  (\ref{deltaEchange}) accounts for switching from the cavity frame to the (Fermi) laboratory frame.
Performing the $z-$integration and recalling (\ref{omegan}) we find
\be
\delta E=\gamma\frac{ L+2 H}{2}\bigg\{\frac{A_\perp}{2(2\pi)^2}\sum_n\int d^2k_\perp \left[k_\perp^2+\left(\frac{n\pi}{L}\right)^2\right]^{1/2}\bigg\}.
\ee
Comparing with (\ref{Ediv}) we see that the part enclosed in curly brackets just represents the flat spacetime Casimir energy (\ref{ecasflat}). Hence we may rewrite $\delta E$ using the flat result (\ref{ecasflat})
\be
\delta E=-\gamma\left(H+\frac{L }{2}\right)A_\perp \frac{\pi^2}{1440 L^3}.
\ee
According to the Fermi observer, the total Casimir energy is
\be\label{Ecastot}
E=-\left[1+\gamma \left(H+\frac{L }{2}\right)\right]A_\perp \frac{\pi^2}{1440 L^3}.
\ee
\subsection{Casimir vacuum energy buoyancy}
Looking at (\ref{Ecastot}) we can identify the quantity $H+L/2$ as the $z$ coordinate of the centre of mass of the vacuum energy stored inside the cavity\footnote{To the present (lowest) order of approximation, we can consider the vacuum energy uniformly distributed inside the cavity, neglecting possible tidal effects. See, e.g., \cite{sorgejj}.}
\be
z=H+\frac{L}{2}.
\ee
Then the Casimir energy reads
\be\label{CE}
E=-A_\perp \frac{\pi^2}{1440 L^3}+\left(-\gamma zA_\perp \frac{\pi^2}{1440 L^3}\right).
\ee
We see that the quantity in round brackets behaves as a {\em gravitational potential energy}
\be\label{Ugrav}
U(z)=-\gamma zA_\perp \frac{\pi^2}{1440 L^3}.
\ee
Notice that $U(z)<0$, implying an upward force acting upon the cavity
\be\label{force}
F_z=-\frac{\partial U}{\partial z}=\gamma A_\perp \frac{\pi^2}{1440 L^3}.
\ee
The present result agrees with that of Calloni {\em et al.} \cite{garufi}, and other similar results appeared in the literature. 

The gravitational interaction on a suspended Casimir cavity gives rise to an upward force ($F_z>0$) which is equal to the weight of the modes of the vacuum that are removed - so to say - from the cavity. Such force can be interpreted as an Archimedes buoyancy force in vacuum. As a consequence, we expect a tiny reduction of the (apparent) weight of a Casimir apparatus.

Notice that - to the present $O(\gamma)$ order of approximation - the total energy (\ref{CE}) of the Casimir apparatus does depend on the reference frame. On the other hand, the measured force is coordinate-independent, thus showing that the gravitational interaction behaves as a truly observable quantity. This is closely related to the gauge-invariance of the gravitational interaction term ${\cal L}_g$ in the lagrangian. 

An interesting issue concerns the possible anisotropy of the Tolman mass-energy (\ref{mphi}).  Indeed, one could wonder whether the gravitational correction to the Casimir mass-energy does depend on the orientation of the Casimir apparatus with respect to the gravitational field.
Actually, the correction to the Casimir mass-energy - eq. (\ref{deltaE}) - involves, to the present lowest approximation order, the {\em flat} spacetime modes of the confined scalar field. Hence, the present correction appears not to be sensitive to the cavity orientation.
However, the issue could become non-trivial, when considering higher approximation orders (or strong gravitational regimes), in which the gravitational distortion upon the field modes has to be taken into account. In this latter case, the apparatus orientation could really matter, due, e.g., to the gravitationally induced tidal effects throughout the cavity (see, e.g., \cite{sorgejj}). All this could eventually lead to anisotropic corrections in the Tolman mass.

Such anisotropy can be readily understood, provided we consider the Casimir apparatus as a {\em composite} system, in which the confined quantum field interacts both with the cavity walls (defining the geometry) and the gravitational field. Changing the apparatus orientation basically changes - so to say - the {\em internal structure} of the system (namely, the allowed quantum field modes), owing to a {\em conspiracy} between the gravitational field and the cavity geometry. The gravitationally induced tidal distortion in the field modes does depend on the orientation of the apparatus, because of its spatial {\em extension}. The latter, in turn, involves the geometry of the cavity, thus enlightening the fundamental role of the physical boundaries. Hence, rotating a Casimir apparatus in the gravitational field causes a {\em change} the physical properties of apparatus itself.

It is well-known that mass anisotropies are not unusual in other contexts, as in condensed matter physics. Unfortunately, in the present case there are severe technical difficulties, mainly due to the necessity of considering a more physical model of cavity (a closed three-dimensional box, rather than a couple of large plates enclosing the quantum field). Apart from some exact solutions describing a little number of selected cavity geometries, only numerical results (and in flat space) are typically at our disposal. Adapting such results to a gravitational background represents indeed an interesting as well stimulating challenge.


\section{Concluding remarks}

In this paper we have reconsidered in detail the influence of a weak gravitational field (like the terrestrial one) on a suspended Casimir apparatus.

As pointed out by Calloni et al. \cite{garufi,avino}, the contribution to the mass due to the quantum field confined to the Casimir cavity does depend also on the spatial components $T^i_{\,\,\,i}$ of the field stress-energy tensor. In spite of this, the force required to balance the weight of the quantum vacuum is related solely to the energy (i.e., the $T^t_{\,\,\,t}$ component of the field stress-energy tensor). Calloni proved that this is indeed a consequence of the red-shift of the force densities acting on the various points inside the cavity.

Here we have shown that the same result can be obtained introducing the notion of quasi-local mass, adapted to the Casimir apparatus. In so doing we have proved that the Tolman mass contribution stemming from the quantum field coincides with the quasi-local value of Casimir energy. Furthermore, our findings agree with those obtained by Calloni and other Authors.

In the weak field limit, the resulting vacuum energy can be decomposed in two parts, one of which represents the gravitational energy due to the interaction between the quantum vacuum and the gravitational field.

It is worthnoting that the total Casimir energy does depend on the adopted reference frame, as it is clearly apparent from (\ref{CE}). This has probably been the main cause of some misunderstanding and contrasting claims appeared in the literature through the years. The results of section 7 and 8 now agree with the recent calculations by Lima {\em et al.} \cite{lima1,lima2} as regards the evaluation of the Casimir energy by means of (\ref{Ecurved}). However, as discussed in section 4, (\ref{Ecurved}) is not eligible as an observable quantity. On the other hand, the interaction with the gravitational field is truly frame-independent, hence {\em observable}, showing that the gravitational field does indeed influence the vacuum energy. Such frame-independence can be traced back to the gauge-invariance of the coupling between matter and gravity.

It is likely that the present approach can be carried on also in presence of soft boundary conditions like, e.g., dissimilar or semitransparent mirrors acting as confining plates \cite{bordag2,barone1,barone2}. Besides, it could be of some interest using the present approach to analyze the behaviour of Casimir energy in presence of different spacetime backgrounds. In that respect, Taub spacetime \cite{taub} represents an example of spacetime with uniform gravitational acceleration, matching the geometry of a {\em standard} parallel-plane Casimir cavity. It maybe that the singular plane, characterizing the Taub solution, plays the role of one of the plates in the Casimir cavity, thus allowing for a single-plate Casimir effect \cite{celestino,saravi}. Those issues will be the subject of future investigation.

Finally, we point out that the success of a much-awaited future experimental measurement of the vacuum weight (as in the {\em Archimedes Experiment}) will surely contribute to shed  light on the present issue, as well as other related open questions concerning, e.g., the Cosmological Constant Problem \cite{caldwell,cerdonio,bengochea}.


\section*{Acknowledgments}
We would like to thank the Referees, whose comments and suggestions were helpful to improve the present manuscript.


\section*{Appendix}

The space-time line element of a weak, static gravitational source $M$  in the Lorentz gauge reads
\be\label{dslorentz}
ds^2=\left(1+2\Phi(x)\right)dt^2-\left(1-2\Phi(x)\right)\left(dx^2+dy^2+dz^2\right),
\ee
where $\Phi(x)=-M/r$ is the newtonian gravitational potential. 
In the weak field limit, we assume $M\ll R$ so that $|\Phi(x)|\ll 1$.
Expanding around the point $r=R$ we may write
\be\label{potexpansion}
\Phi(x)=-\frac{M}{R}+\frac{M}{R^2}z=\Phi_0+\gamma z+O(\gamma^2),
\ee
where $\Phi_0=-M/R$ and $\gamma =M/R^2$. So
\be\label{dslorentz2}
ds^2=\left(1+2\Phi_0+2\gamma z\right)dt^2-\left(1-2\Phi_0-2\gamma z\right)\left(dx^2+dy^2+dz^2\right).
\ee
Let us now perform the following coordinate transformation
\be\label{coordtrans}
\left\{
	\begin{array}{lll}
	t&=&t'	\\
	x&=&x'+\Phi_0x'+\gamma x'z'\\
	y&=&y'+\Phi_0y'+\gamma y'z'\\
	z&=&z'+\Phi_0z'+\frac{1}{2}\gamma z'^2-\frac{1}{2}\gamma x'^2-\frac{1}{2}\gamma y'^2.	
	\end{array}
	\right.
\ee
Differentiating (\ref{coordtrans}) and substituting in (\ref{dslorentz2}) we find, to the $O(\gamma)$ order
\be\label{cfermi1}
ds^2=\left(1+2\Phi_0+2\gamma z'\right)dt'^2-\left(dx'^2+dy'^2+dz'^2\right).
\ee
We may also remove the constant part $2\Phi_0$ by means of a further coordinate change, namely $t'=(1-\Phi_0)t''$. Dropping the primes at the end of calculations, we obtain
\be\label{cfermi2}
ds^2=\left(1+2\gamma z\right)dt^2-dx^2-dy^2-dz^2,
\ee
namely the metric in the Fermi coordinates, see (\ref{ds}).

We point out that the underlying physics is not altered by the above transformations, thanks to the gauge-invariance of the gravitational interaction term ${\cal L}_g=-\frac{1}{2}h_{\mu\nu}T_{(0)}^{\mu\nu}$ appearing in (\ref{Lexpand2}).

To see this in some detail, consider the following gauge transformation of the coordinates
\be\label{gaugecoordchange}
x^\mu\rightarrow x'^\mu=x^\mu+\epsilon^\mu(x^\nu),
\ee
where $\partial\epsilon^\mu/\partial x^\nu$ is at most of the same order of magnitude as $h_{\mu\nu}$.
The change in the gravitational potentials is
\be\label{hchange}
h_{\mu\nu}\rightarrow h'_{\mu\nu}= h_{\mu\nu}-\partial_\mu\epsilon_\nu -\partial_\nu \epsilon_\mu.
\ee
The corresponding change $\delta S$ in the matter action (\ref{S}), due to the gravitational interaction, reads (to the $O(\gamma)$ order) 
\bea\label{gaugeS}
\delta S=\int_\Omega d^4x\,\delta{\cal L}_g=\int_\Omega d^4x\left(\partial_\mu\epsilon_\nu\right) T_{(0)}^{\mu\nu}\nonumber\\
=\int_\Omega d^4x\,\partial_\mu\left(\epsilon_\nu T_{(0)}^{\mu\nu}\right)-\int_\Omega d^4x\,\epsilon_\nu\partial_\mu T_{(0)}^{\mu\nu}=0.
\eea
Actually, the first integral is zero, since it can be converted to a surface integral vanishing on the boundary $\partial\Omega$, where $T_{(0)}^{\mu\nu}\rightarrow 0$. The last integral is zero, due to the continuity equation $\partial_\mu T_{(0)}^{\mu\nu}=0$. Hence, the change of the matter action induced by the gauge transformation (\ref{gaugecoordchange}) is $\delta S=0$. This proves the gauge-invariance of $S$ with respect to the coordinate transformation (\ref{gaugecoordchange}).


\end{document}